\documentclass[epj,draft]{svjour}
\usepackage{amsfonts}
\usepackage{amssymb}
\usepackage{graphicx}
\usepackage{bm,amsmath}

\begin{document}
\title{Theoretical overview of atomic parity violation}
\subtitle{Recent developments and challenges}
\author{Andrei Derevianko\inst{1} \and Sergey G. Porsev\inst{2}
}                     
\offprints{andrei@unr.edu}          
\institute{
Physics Department, University of
Nevada, Reno, Nevada  89557, USA
 \and Petersburg Nuclear Physics Institute, Leningrad
district, Gatchina, 188300, Russia}
\date{Received: date / Revised version: date}
%
\abstract{
Recent advances in interpreting
the most accurate to-date  measurement of atomic parity violation in Cs
are reviewed. The inferred nuclear weak charge,
$Q_W(^{133}\mathrm{Cs}) =  - 72.65(28)_\mathrm{expt} (36)_\mathrm{theor}$,
 agrees with the prediction of
the standard model at $1\,\sigma$ level. Further improved interpretation
is limited by an accuracy of solving
basic correlation problem of atomic structure. We report on our
progress in solving this problem within the relativistic coupled-cluster formalism.
We 
include single-, double- and triple- electronic excitations
in the coupled-cluster expansion.
Numerical results for energies,
electric-dipole matrix elements, and hyperfine-structure constants of Cs are
presented.
} 
\maketitle
\section{Introduction}
\label{intro}

This year, year of 2006, marks 50 years of the fall of parity conservation, one of the
hallmark discoveries of the 20th century physics.
The field has started  with
the seminal Lee and Yang paper~\cite{LeeYan56} and discovery of
parity violation in the nuclear $\beta$-decay~\cite{WuAmbHay57}. Soon
after this discovery, Zel'dovich~\cite{Zel59} contemplated  possibility of
observing parity-nonconserving (PNC) signal in atoms. He
concluded that the effect was too small to be of experimental
significance. In 1974, however,  M.-A. Bouchiat and C.
Bouchiat~\cite{BouBou74} realized that the PNC is amplified in {\em heavy}
atoms. They showed that the relevant PNC amplitude scales steeply
with the nuclear charge $Z$.
In atomic physics, the first P-violating signal has been observed in
1978 by Barkov and Zolotorev~\cite{BarZol78} in Bi atom.
Over the following decades the experiments were refined, with
PNC signal observed in several atoms. 
So far the most accurate measurement has been carried out
in $^{133}$Cs by Boulder group~\cite{WooBenCho97}.

Rich history of atomic parity violation is examined in a
number of review articles, e.g., Refs.~\cite{BouBou97,GinFla04} and a book~\cite{Khr91}.
Here, due to the space limitation,
we restrict our attention to  recent developments in interpreting
P-violation in Cs atom
and report on the progress towards more accurate interpretation
of atomic PNC in this atom.

The PNC measurements are
interpreted in terms of the weak nuclear charge $Q_W$, which
quantifies the strength of the electroweak coupling between atomic
electrons and quarks in the nucleus. The relation between $Q_W$ and
the PNC amplitude, $E_\mathrm{PNC}$, can be represented as
\begin{equation}
E_\mathrm{PNC} = k \, Q_W \, ,
\label{Eq:kDef}
\end{equation}
where $k$ is an atomic-structure factor. Apparently, the
interpretation requires atomic-structure calculations of $k$ with an
accuracy that matches the experimental uncertainty in
$E_\mathrm{PNC}$. In particular, in Cs, the present theoretical
uncertainty dominates over experimental error bar resulting in an
overall  0.6\% uncertainty in the inferred value of
$Q_W(^{133}\mathrm{Cs})$.

The underlying theory of the electroweak interactions
is well established and
its predictions have been verified in a variety of experiments. Much
of the present efforts are driven by searches for ``new physics''
beyond the standard model (SM). We may  distinguish between two
approaches to such searches: low- and high-energy experiments.
Atomic parity violation probes the low-energy
electroweak sector  of the SM. While the relevant
momentum transfer is just $\sim$ 30 MeV, the exquisite accuracy of
the interpretation constrains new physics at much higher energies.
Indeed, following arguments of Ref.~\cite{Ram99}, we find that the
present 0.6\% determination of the $Q_W(^{133}\mathrm{Cs})$ probes
the new physics at a mass scale of
\[
\left\{
8 \, \sqrt{2} \, \pi \frac{1}{G_F} \, \left|\frac{Q_W}{\delta Q_W }\right|
\right\}^{1/2} \approx 20\,
\mathrm{TeV} \, ,
\]
where $G_F$ is the Fermi constant. Apparently by reducing the
uncertainty in the determination of the weak charge, $|\delta Q_W|$,
we may access even higher mass scales. Such a reduction in the
uncertainty of $Q_W$ is the goal of our present efforts outlined in Sec.~\ref{Sec:MBPT}.

While discussing the impact of atomic PNC on particle physics, it is
worth noting that colliders are blind to certain ``new physics''
scenarios~\cite{Ram99}. For example, atomic PNC is uniquely
sensitive to extra  Z bosons 
predicted in grand
unified theories, technicolor models, SUSY, and string theories.
Generally, atomic PNC is one of a few probes of electroweak coupling
below the $Z$-pole. Competing and at the same time complementary
experiments include determination of the weak charge of electron at
SLAC~\cite{AntArnArr05} and $\nu$-nucleus deep inelastic
scattering by the NuTeV collaboration~\cite{ZelMcFAda02}. With an
upgrade at Jefferson Lab, an even more accurate constraint on
electroweak coupling is expected to come from a
P-violating $e-p$ scattering experiment (Qweak
collaboration)~\cite{QweakWWW}. Still, the atomic PNC will continue
to play an important complementary role, as the atomic PNC is
sensitive to electron-neutron couplings, while the Qweak experiment
will probe electron-proton coupling, i.e., a different linear
combination in terms of the up and down quark-electron couplings.

\section{Deviation from the Standard Model and recent theoretical
progress} Parity violation in Cs has been a subject of a recent
controversy. To infer the value of the weak charge, one needs to
combine measurements with atomic-structure calculations. The
uncertainty in the value of the inferred $Q_W$ is usually determined
by summing experimental and theoretical uncertainties in quadrature,
\begin{equation}
\sigma = \left\{\sigma^2_\mathrm{expt}  + \sigma^2_\mathrm{theor}
\right\}^{1/2} \, . \label{Eq:sigQuadrature}
\end{equation}
In 1997, the most accurate to date (0.35\%) measurement of PNC has been
carried out by Boulder group~\cite{WooBenCho97}. At that time, the
accuracy of  calculations~\cite{DzuFlaSus89,BluJohSap90} has
been estimated to be 1\%. In 1999, Bennet and Wieman~\cite{BenWie99},
in light of new lifetime data which improved theory-experiment
agreement, have reduced the theoretical error bar to 0.4\%. While
compared  with the prediction of the SM, the resulting
value of $Q_W$ differed by 2.5 $\sigma$, one of the largest
deviations in the recent history. The deviation has prompted a
substantial interest from particle physics community (see, e.g.,
\cite{Ram99,CasCurDom99,Ros00,Ros02}). At the same time the
reduced theoretical 
uncertainty raised the questions whether some "small"
sub-1\%  atomic-structure effects could be the reason for the
deviation.

\begin{table*}[h]
\begin{center}
\centering
\begin{tabular}[c]{lcl}
\hline\hline
\begin{tabular}{l}
Deviation (1999)\\
Bennet \& Wieman~\cite{BenWie99}
\end{tabular}
 & $2.5 \sigma$  &
\begin{tabular}{l}
Based on calculations by \\
 Dzuba {\em et al.} (1989)~\cite{DzuFlaSus89} and \\
Blundell {\em et al.} (1990)~\cite{BluJohSap90}.
\end{tabular}
 \\
\hline \multicolumn{1}{c}{Theoretical correction}& Pull  &
\multicolumn{1}{l}{\, \, Reference} \\
\hline
\begin{tabular}{l}
Breit interaction
\end{tabular}
&  $-1.2 \sigma$  &
\begin{tabular}{l}
 Derevianko (2000)~\cite{Der00}; \\
 Dzuba {\em et al.} (2001)~\cite{DzuHarJoh01};\\
 Kozlov {\em et al.} (2001)~\cite{KozPorTup01}. \\
\end{tabular}
\\[3ex]

\begin{tabular}{l}
Vacuum polarization ( $+ 0.8 \sigma$)\\
Vertex/self-energy ( $-1.3 \sigma$) \\
\, \\
\, \\
\, \\
\, \\
\end{tabular}
& $-0.5 \sigma$ &
\begin{tabular}{l}
Johnson {\em et al.} (2002)~\cite{JohBedSof01}; \\
Kuchiev \&
Flambaum (2002)~\cite{KucFla02}; \\
Milstein {\em et al.} (2002)~\cite{MilSusTer02,MilSusTer03a};\\
Sapirstein {\em et al.} (2003)~\cite{SapPacVei03}; \\
Flambaum  \& Ginges (2005)~\cite{FlaGin05};\\
Shabaev {\em et al.} (2005)~\cite{ShaPacTup05}.
\end{tabular}
\\[2ex]
\begin{tabular}{l}
Neutron skin
\end{tabular}
& $-0.4 \sigma$ &
\begin{tabular}{l}
Derevianko (2002)~\cite{Der02}
\end{tabular}
\\[2ex]
\begin{tabular}{l}
Updated correlated $E_\mathrm{PNC}$ and $\beta$
\end{tabular}
& +0.7$\sigma$&
\begin{tabular}{l}
Dzuba {\em et al.}  (2002)~\cite{DzuFlaGin02}.
\end{tabular}
\\[2ex]
\begin{tabular}{l}
$e-e$ P-violating  interaction, \\
Renormalization $Q\rightarrow 0$, \\
Giant nuclear resonance  \\
\end{tabular}
& $-0.08 \sigma$ &
\begin{tabular}{l}
Sushkov \& Flambaum (1978) \\
Milstein   {\em et al.} (2002)~\cite{MilSusTer02}
\end{tabular}\\
\hline
Total deviation (agreement) & $1  \sigma$ & \\
\hline\hline
\end{tabular}
\caption{ Present status of the  $2.5\sigma$ deviation of inferred
$Q_W(^{133}\mathrm{Cs})$  from the prediction of the Standard Model.
 In the bulk of the Table, we summarize the recent theoretical progress on
 computing ``small'' sub-1\% corrections to the $6s-7s$ PNC amplitude. These corrections
 directly affect the inferred value of  the weak charge of Cs. For illustrative purposes
 all the corrections (pulls) are tabulated in terms of the original
 Bennet and Wieman~\cite{BenWie99} uncertainty,
 $\sigma = \left\{\sigma^2_\mathrm{expt}  + \sigma^2_\mathrm{theor} \right\}^{1/2} \approx 0.5\%$. }
\label{Tab:recentQCshistory}
\end{center}
\end{table*}

Over the last few years, there has
been an important progress in understanding ``small''
corrections, such as Breit, radiative, and neutron-skin corrections. These
advances are summarized in Table~\ref{Tab:recentQCshistory}.

Taking into account the tabulated ``small'' corrections, revised value of vector
transition polarizability and somewhat improved  value of $E_\mathrm{PNC}$~\cite{DzuFlaGin02},
we arrive at the following revised value of the weak charge
(the assigned theoretical 
uncertainty of 0.5\%
is consistent with an estimate of Ref.~\cite{DzuFlaGin02} and
the central value with that of Ref.~\cite{ShaPacTup05})
\begin{equation}
Q_W(^{133}\mathrm{Cs}) =  - 72.65(28)_\mathrm{expt} (36)_\mathrm{theor}
\end{equation}
This value agrees with the SM value~\cite{Ros02} of $-73.19(13).$ at $ 1\,\sigma$ level.

 The present theoretical
uncertainty in PNC calculations is estimated to be 0.5\%. Since the
overall error is obtained with quadrature~(\ref{Eq:sigQuadrature}),
reducing $\sigma_\mathrm{theor}$ below the experimental uncertainty
of 0.35\% will either reveal new physics or set important
constraints on competing extensions to the SM. In addition, it will
set a stage for further experimental efforts, such as those by Paris
group~\cite{GueChaJac03,GueLinBou05}. Other experimental efforts
which will benefit from the improved calculations are PNC
measurements with  Ba$^+$\cite{For93,KoeSchNag03},
Tl~\cite{VetMeeMaj95}, and Fr~\cite{GomOroSpr06}.

\section{High-accuracy atomic-structure calculations}
\label{Sec:MBPT}

The present accuracy in the determination of
the weak charge of Cs is limited by the accuracy of solving the
basic correlation problem. The many-body problem is hard. Even
classically, the three-body problem can not be solved in closed
form. While adequate numerical approaches (e.g., configuration
interaction and variational methods) were developed for few-electron
atoms, accurate solution of the many-body problem for atoms with
many electrons still remains a challenge.

In a particular case of the PNC problem in Cs, we need to evaluate
the PNC amplitude
for the $6S_{1/2} \rightarrow 7S_{1/2}$ transition
\begin{eqnarray}
\lefteqn{E_\mathrm{PNC} = \sum_{n}
\frac{\langle 7S|D|nP_{1/2}\rangle  \langle nP_{1/2} |H_{W}|6S\rangle
}{E_{6S}-E_{nP_{1/2}}}  } \nonumber \\  &+ &
\sum_{n}
\frac{\langle 7S|H_{W}|nP_{1/2}\rangle  \langle nP_{1/2} |D|6S\rangle
}{E_{7S}-E_{nP_{1/2}}}
\, .
\label{Eq:EPNC}
\end{eqnarray}
Here $D$ and $H_{\rm W}$ are electric-dipole and weak interaction matrix elements,
and $E_{i}$ are atomic energy levels. The effective weak interaction, averaged over
quarks is accumulated inside the nucleus
\begin{equation}
 H_{\rm W} = -\frac{G_F}{\sqrt{8}} \, Q_W \,  \gamma_5 \,
 \rho ({\bf r}) \,,
 \label{Eq:HW}
\end{equation}
where $\rho ({\bf r})$ is the  neutron-density
distribution. One could easily recast
Eqs.~(\ref{Eq:EPNC}) and (\ref{Eq:HW}) in terms of the structure factor $k$ of Eq.~(\ref{Eq:kDef}).

We wish to 
evaluate accurately the sum~(\ref{Eq:EPNC}).
Let us comment on pre-requisites for such calculations.
(i) Due to the particular
characters of the involved operators in summation (\ref{Eq:EPNC}), the approximate wave-functions
have to reproduce the short-range (close to the nucleus) and
long-range atomic properties simultaneously. The accurate short-range
description requires {\em ab initio} relativistic approach, as the
ratio of velocity of electron to the speed of light near the nucleus
is $\sim \alpha Z \sim 0.4$ for Cs. (ii) With respect to the accuracy,
we notice that simple Dirac-Hartree-Fock (DHF) calculations
for the hyperfine-structure (HFS) constant of the ground state are about
50\% off from the experimental value.
This constant describes strength of coupling of the electron to the
nuclear magnetic moment and its short-distance nature mimics 
behavior of the weak matrix elements. Therefore we arrive at a conclusion
that so-called correlation corrections (those beyond the DHF approximation)
have to be addressed.
We treat the correlations within the systematic and
successful methods of  many-body perturbation theory.

How do we estimate the accuracy of calculating $E_\mathrm{PNC}$?
Ultra-precise experimental data are available for Cs. These data have accuracy better
than our anticipated theoretical 
 uncertainty thus allowing us to calibrate our calculations.
Atomic energies are known to many significant 
 figures.
Ultra-precise (0.01\% accurate) value of lifetime of the $6P_{3/2}$ state of Cs has been
recently inferred from photoassociation spectroscopy~\cite{AmiDulGut02,BouCruDul06}
with ultracold atoms.
The HFS constants for Cs are also measured to a good
accuracy (the  HFS constant $A_{6S}$ of $^{133}$Cs is
known {\em exactly} by the definition of the second).

To reiterate, to further improve interpretation of atomic parity
violating signals, we need a high-accuracy {\em ab initio} relativistic
many-body method capable of reaching { the} accuracy level of 0.1\%
for Cs.

\section{Relativistic coupled-cluster method}
Many-body perturbation theory (MBPT) provides a systematic way of
treating correlation corrections, the critical issue for
an improved interpretation of atomic PNC. First we review
main ideas of MBPT and the all-order coupled-cluster (CC) method.
Then we describe our present CC-inspired computational
scheme and illustrate it with numerical results for
basic atomic properties of Cs atom.

\subsection{Generalities}

In MBPT the atomic
Hamiltonian  is partitioned as
\begin{eqnarray*}
 H &=& \left( \sum_{i}\,h_{\rm nuc}(\mathbf{r}_i) +
\sum_{i}\, U_{\rm DHF}(\mathbf{r}_i) \right) + \\
&& \left(
\frac{1}{2}\sum_{i \neq j}\frac{1}{r_{ij}} -\sum_{i}\,U_{\rm
DHF}(\mathbf{r}_i) \right) \,,
\end{eqnarray*}
where $h_{\rm nuc}$ includes the kinetic energy of an electron and
its interaction with the nucleus, $U_{\rm DHF}$ is the
DHF potential, and the last term represents the
residual Coulomb interaction between electrons. In MBPT the first
part of the Hamiltonian is treated as the lowest-order Hamiltonian
$H_0$ and the residual Coulomb interaction as a perturbation.
The perturbative expansion
is built in powers of residual interaction and the derivations
typically involve methods of second quantization and diagrammatic
techniques.

One of the mainstays of practical applications of MBPT is an
assumption of convergence of series in powers of the residual
interaction. Sometimes the convergence is poor and then one sums
certain classes of diagrams  to ``all orders'' using iterative
techniques. One of the most popular all-order methods is the
coupled-cluster (CC)  formalism~\cite{CoeKum60,Ciz66}. It is widely
employed in atomic and nuclear physics, and quantum
chemistry~\cite{BisKum87}. The  {\em relativistic} atomic-structure
CC-type calculations were carried out, for example, in
Refs.~\cite{BluJohLiu89,BluJohSap91,EliKalIsh94,AvgBec98,SafDerJoh98,SafJohDer99}.

The Hamiltonian in the second quantization (based on the DHF basis) reads
\begin{eqnarray}
 H &=&  H_0 + G \nonumber \\
 &=& \sum_{i} \varepsilon_i \left\{ a_i^\dagger a_i \right\} +
\frac{1}{2} \sum_{ijkl} g_{ijkl} \left\{ a^\dagger_i a^\dagger_j a_l
a_k \right\} \, , \label{Eq:SecQuantH}
\end{eqnarray}
where $H_0$ is the lowest-order Hamiltonian and the residual Coulomb
interaction $G$ is treated as a perturbation. The operators $a_{i}$
and $a_{i}^\dagger$ are annihilation and creation operators, and
$\left\{... \right\}$ stands for a normal product of operators with
respect to the core quasivacuum state $|0_c\rangle$.
In the lowest order the atomic wave function with the valence
electron in an orbital $v$ reads $|\Psi_v^{(0)} \rangle =
a^\dagger_v | 0_c \rangle$. For example, $v$ can represent the ground
state $6s$ orbital of Cs atom.
Formally we can introduce  a wave operator
$\Omega$ that produces
the exact many-body wave function, $|\Psi_v\rangle = \Omega\, |\Psi_v^{(0)}\rangle$.

Central to the CC method is realization that the 
{ wave operator} can be represented via exponential ansatz~\cite{LinMor86}
\begin{equation}
 \Omega = \left\{\exp(K)\right\} =
 1 + K + \frac{1}{2!} \left\{K^2\right\} + \ldots \, .
 \label{Eq:CCOmega}
\end{equation}
The operator $K$ can be compellingly separated into cluster operators combining
simultaneous excitations of core and valence electrons from the
reference state $|\Psi_v^{(0)}\rangle$ to all orders of MBPT,
\begin{equation}
K = S + D + T + \cdots,
\end{equation}
i.e., $K$ is separated into single ($S$), double ($D$), triple
($T$), and higher-rank excitations. For the univalent systems it is
convenient to subdivide cluster operators into  core and
valence classes
\begin{equation}
 K = K_c + K_v \,.
\end{equation}
Clusters $K_c$ involve excitations from the core orbitals only, while
$K_v$ describe simultaneous excitations of the core and valence
electrons. Then $S=S_c + S_v$, $D=D_c+D_v$, etc.
The cluster operators satisfy so-called Bloch equations~\cite{LinMor86}.

While the CC formulation is exact, in practice
the full cluster operator $K$ is truncated at a
certain level of excitations. If we restrict it to single and
double excitations
\begin{equation}
K \equiv K^{({\rm SD})} \approx S_c + D_c + S_v + D_v
\end{equation}
we arrive at the widely employed coupled-cluster single double (CCSD) method.

{\em Linearized} version of the CCSD method
was  employed by the
Notre Dame group for high-accuracy calculations of various atomic
properties~\cite{BluJohLiu89,BluJohSap91,SafDerJoh98,SafJohDer99}.
In this approximation, one discards nonlinear terms in the
expansion of exponent in Eq.~(\ref{Eq:CCOmega}),
$\Omega \approx 1 + K^{({\rm SD})}$. We will refer
to this approximation as the singles-doubles (SD) method.
The resulting SD equations are written out in
Ref.~\cite{BluJohLiu89}. A typical {\em ab initio} accuracy attained
for properties of heavy alkali-metal atoms is at the level of 1\%.

Since the present goal is to reduce theoretical uncertainties
to the level of 0.1-0.2\% we have to go
beyond the SD approach. A systematic step in improving the SD method
would be an additional inclusion of triple and nonlinear double
excitations. However, considering the present state of available
computational power, the full incorporation of triples
(specifically, core triples) seems to be unmanagable for heavy
atoms. For instance, for Cs storing and manipulating core triple
amplitudes  would require $\sim$ 100 Gb of  memory.

To motivate next-generation formalism, we
have explicitly computed 
{ 1648} fourth-order diagrams
for matrix elements that appear due to triple excitations
and non-linear terms (i.e., those omitted in the SD method)~\cite{DerEmm02,CanDer04}.
We observe from numerical results for
electric-dipole matrix elements in Na~\cite{CanDer04}
and Cs~\cite{DerPor05} that the contributions from {\em
valence} triples $T_v$  and nonlinear
doubles $D_{nl}$  are much larger than
those from {\em core} triples $T_c$.
This leads to our present level of
approximation: we discard core triples { and core nonlinear terms}
and incorporate the {\em valence triples} and {\em
valence nonlinear terms} into the SD formalism. The resulting
approximation will be referred to as CCSDvT method.

\subsection{Driving equations in the CCSDvT approximation}

Below we write down the CC equations for cluster amplitudes
in the CCSDvT approximation. Here we present topological
structure of the equations only.
A detailed tabulation of the formulas can be found in  our
paper~\cite{PorDer06Na}.
The equations in the SD approximation are presented in
explicit form in Ref.~\cite{BluJohLiu89}. The CCSDvT equations for the core cluster
amplitudes $S_c$ and $D_c$ are the same as in the SD approximation.

For valence triple amplitudes we obtain symbolically
\begin{equation}
- [H_0, T_v]  + \delta E_{v} T_v
 \approx
 T_v[D_c]+ T_v[D_v]  \,.
\end{equation}
Here $[H_0, T_v]$ is a commutator, and $\delta E_v$ is the correlation valence
energy defined as
\begin{equation}
\delta E_v = \delta E_\mathrm{SD} + \delta E_\mathrm{CC} + \delta
E_\mathrm{vT} \,, \label{Eq:delta_Ev} \,
\end{equation}
where correction $\delta E_\mathrm{SD}$ is obtained within
the SD approach, correction $\delta E_\mathrm{CC}$ comes from
nonlinear CC contributions and $\delta E_\mathrm{vT}$ is due to
valence triples.
{ Contributions $T_v[D_c]$ and $T_v[D_v]$ denote the effect of core
and valence doubles on valence triples, respectively.
At present we include only these effects}
omitting the effect of valence and core triples on valence triples
($T_v[T_v]$ and $T_v[T_c]$) and nonlinear CC contributions.
These are higher-order effects which computationally are much more
demanding.


The topological structure of the valence singles equation is
\begin{eqnarray}
\lefteqn{- [H_0, S_v]  + \delta E_{v} S_v \approx  {\rm SD} + }\nonumber \\
&&  S_v[S_c \otimes S_v]  + S_v[S_c \otimes S_c] + \nonumber\\
&&  S_v[S_c \otimes D_v]  +  S_v[S_v \otimes D_c] + S_v[T_v] \, .
\label{Eq:Sv}
\end{eqnarray}
Here $S_v[S_c \otimes S_v]$ stands for a contribution from the
excitations of core and valence electrons resulting from  a
product of clusters $S_c$ and $S_v$. All other
terms are defined in a similar fashion.

Finally, equation for valence doubles can be symbolically represented as
\begin{eqnarray}
\lefteqn{- [H_0, D_v]  + \delta E_{v} D_v \approx  {\rm SD} + }\nonumber \\
&& D_v[S_c \otimes S_v] + D_v[S_c \otimes S_c] +  \nonumber\\
&& D_v[S_c \otimes D_v] + D_v[S_v \otimes D_c] + D_v[S_c \otimes D_c] + \label{Eq:Dv} \\
&& D_v[D_c \otimes D_v] + D_v[S_c \otimes T_v] + D_v[S_v \otimes T_c] + D_v[T_v]
\, . \nonumber
\end{eqnarray}

Solution of the above equations provides us with the cluster amplitudes and
correlation energies. Numerical results for the energies will be presented
in Section~\ref{Sec:numerics}. At this point,
with the obtained wave functions we proceed to evaluating
matrix elements.

\subsection{Matrix elements}

The SD method has already proven to be
successful in  calculations of various atomic
properties. For heavy alkali-metal atoms the attained level of
agreement with experimental data for the hyperfine constants is at
5\% and the accuracy of a similar calculation for the
electric-dipole amplitudes is 0.5\% (see, e.g.,~\cite{SafJohDer99}).
At the same time the accuracy required for our goals should be at
the level of 0.1-0.2\%. In order to improve the overall accuracy we
develop the technique of relativistic calculations of matrix
elements beyond the SD approach.

Given two computed CCSDvT wave functions, we may evaluate matrix elements of
{ one-electron} operator $Z$ as
\begin{equation}
 Z_{wv} =
 \frac{ \langle \Psi_w | \sum_{ij} z_{ij} \, a^\dagger_i a_j | \Psi_v \rangle }
 { \sqrt{ \langle \Psi_w | \Psi_w \rangle \langle \Psi_v | \Psi_v \rangle  }}
 \label{Eq:Zmel} \, .
\end{equation}
The explicit expressions are given in Ref.~\cite{PorDer06Na}. Compared to
the SD approximation, we include contribution of valence triples 
{ $T_v$. They} contribute both directly via explicit contributions to matrix element
formula and indirectly through modification of the SD amplitudes.

It is worth pointing out, that the importance of the valence triples
has been realized earlier by the Notre Dame group~\cite{BluJohSap91}. They have shown
that at the SD level, the error for the HFS constants is as large
as 5\% for Cs. To rectify this problem, they  proposed and implemented
a scheme that approximates the effect $S_v[T_v]$, i.e., effect of
valence triples on valence singles.
While improving agreement for the HFS constants, their approximation
leads to poorer agreement (compared to SD method) for the dipole matrix elements.
The advantage of the Notre  Dame scheme is that
it avoided  expensive storing of triple excitations.
{ Due to improved computational resources, we are able to store triples.
Accounting for the triples in a rigorous fashion leads to
a better agreement between theory and experiment.}

Compared to the Notre Dame approximation we also include 
dressing of matrix elements based on the CC ansatz. The idea of our method\cite{DerPor05} is
as follows. When the CC exponent
is expanded in Eq.~(\ref{Eq:Zmel}), we encounter infinite number of terms.
We devised a method of partial summation (dressing) of the resulting
series. Our formalism is built upon an expansion of the product of
cluster amplitudes into a sum of $n$-body insertions. We considered
two types of insertions: particle (hole) line insertion (line
``dressing'') and two-particle (two-hole)
random-phase-approximation-like  insertion. We demonstrated how
to ``dress'' these insertions and formulated iterative equations.

Another formal improvement over Notre Dame calculations
comes from including the CC nonlinear terms in the equations
for valence singles~(\ref{Eq:Sv}) and doubles~(\ref{Eq:Dv}). We also
 include  contribution of
{ the} core triples to matrix elements from direct fourth-order calculation.

\subsection{Numerical results}
\label{Sec:numerics}
Our developed numerical CCSDvT code is an
extension of the relativistic SD code~\cite{SafDerJoh98} which
employs B-spline basis set. This basis  numerically approximates
complete set  of single-particle atomic states. Here we use 35 out
of 40 positive-energy basis functions. Basis functions with $l_{\rm
max} \le 5$ are used for singles and doubles. For triples we employ
a more limited set of basis functions with $l_{\rm max}(T_v) \le 4$.
Excitations from core sub-shells [4$s$,...,5$p$] are included in the
calculations of triples while excitations from sub-shells
[1$s$,...,3$d$] are discarded.

\begin{table}[h]
\caption{Contributions to removal energies of $6s$, $6p_{1/2}$, and
$6p_{3/2}$ states for Cs in cm$^{-1}$ in various approximations.
$\delta E^{\rm tot}_{\rm extrapolated}$ correction is obtained by
computing SD properties with  increasingly larger basis sets and
interpolating them to $l=\infty$~\cite{Saf00}. A comparison with
experimental values is presented in the lower panel. }
\label{Tab:Cs_E}
\begin{center}
\begin{tabular}{lrrr}
\hline \hline
\smallskip
& \multicolumn{1}{c}{$6s$}
& \multicolumn{1}{c}{$6p_{1/2}$}
& \multicolumn{1}{c}{$6p_{3/2}$}\\
\hline
\smallskip
 $E_{\rm DHF}$              & 27954   & 18790    & 18389   \\
 $\delta E_{\rm SD}$        &  3869   &  1611    &  1623  \\
$\delta E_{\rm CCSDvT}$     &   3350  &   1387   &   1220  \\
\smallskip
$E^{\rm tot}_{\rm CCSDvT}$  &  31304  &  20178   &  19608 \\
QED \footnotemark[1]
                            &     18  &     -0.4 &      0 \\
\smallskip
$\delta E^{\rm tot}_{\rm extrapolated}$
                            &     30  &     20   &     20 \\
\hline
$\rm E^{\rm tot}_{\rm final}$&  31352 &  20198   &  19628 \\
{\rm $E_{\rm experim}$} \footnotemark[2]
                            &  31407  &  20228   &  19675  \\
\hline \hline
\end{tabular}
\end{center}
\footnotemark[1]{Reference~\cite{FlaGin05}};
\footnotemark[2]{Reference~\cite{Moo58}}.
\end{table}

Computed removal energies of $6s$, $6p_{1/2}$, and $6p_{3/2}$ states
of atomic cesium are presented in Table~\ref{Tab:Cs_E}. The dominant
contribution to the energies comes from the DHF values. The
remaining (correlation) contribution is given by
Eq.~(\ref{Eq:delta_Ev}). We computed this correlation correction in
SD and  CCSDvT approximations. As it
follows from the table the agreement with experiment is at the level
of 0.1-0.2\% for all considered states. We anticipate that including
other corrections missed at this stage (e.g., nonlinear corrections
to core amplitudes and core triples) can further improve the
agreement with the experimental results.

With the computed wave functions of the $6s$, $6p_{1/2}$ and
$6p_{3/2}$ states we  determine matrix elements. Numerical results
for magnetic-dipole hyperfine-structure constants $A$ and
electric-dipole transition amplitudes are presented in
Table~\ref{Tab:Cs_MEs}. This Table is organized as follows. First we
list the DHF and SD values, and the differences between CCSDvT and
SD values, $\Delta(\mathrm{CCSDvT}) = \mathrm{CCSDvT} -
\mathrm{SD}$.
We base our final {\em ab initio} results on the most sophisticated
CCSDvT values. These { values} also include all-order dressing,
and corrections due to core triples, computed in the the fourth
order of MBPT. QED corrections are included where available. The
results for the HFS constants include finite-nuclear size
(Born-Weisskopf) effect.

We find an excellent, 0.1\%-level, agreement for dipole matrix elements and
the HFS constant of the ground state. The agreement for the
the HFS constant of the $6p_{1/2}$ is only at 1\% level. We are presently
working on testing sensitivity of this constant
to higher-order effects.

We would like to emphasize that presently the correlation corrections
at the level of a few 0.1\% are comparable to radiative corrections.
In this regard it would be useful to carefully compute them
to unmask the remaining many-body effects.

\section{Summary and outlook}
Atomic parity violation plays an important role in testing low-energy electroweak
sector of standard model. Interpretation of  experiments in terms
of nuclear weak charge requires  calculations challenging the capabilities
of modern atomic theory.
Over the last few years, we have 
witnessed a substantial advance in evaluating  corrections to
parity violating amplitudes in heavy atoms. These small (sub-1\%), but important
 corrections include Breit, radiative (vacuum polarization, self-energy, and
vertex) and neutron skin corrections. As a result of this progress,
the most accurate to date measurement of atomic parity violation in
Cs has been brought into substantial agreement with the prediction
of the standard model.

Presently the  theoretical interpretation is clouded by
uncertainties in solving the basic correlation problem of atomic
structure. In this paper we outlined our next-gen\-eration many-body
formalism for solving this problem. We tested our
coupled-cluster-inspired method by computing basic atomic properties
of Cs atom.  All the computed properties are important for
quantifying accuracy of the calculations of parity-violating
amplitudes. We find an agreement at 0.1\% for the ground state
 hyperfine structure constant, 
{\ E1 transition amplitudes}, and energies. However, a relatively
poor 1\% agreement of the  HFS constant $A$ for the
$6P_{1/2}$ {state} with experiment requires further improvements
of the method. The advantage of the employed coupled-cluster method
is that it allows for such systematic improvements. It is anticipated
that the further theoretical progress will refine constraints on  new physics
beyond the standard model and enable next round of experimental studies.

\begin{table*}[h]
\caption{Magnetic-dipole hyperfine structure constants $A$ (in MHz)
and matrix elements of electric dipole moment (in a.u.) for
$^{133}$Cs. Results of calculations and comparison with experimental
values are presented. See text for the explanation of entries.
\label{Tab:Cs_MEs} }
\begin{center}
\begin{tabular}{lr@{.}lr@{.}lr@{.}lr@{.}l}
\hline \hline
 & \multicolumn{2}{c}{$ A(6s) $}     & \multicolumn{2}{c}{$ A(6p_{1/2})$}
 & \multicolumn{2}{c}{$\langle 6p_{1/2} ||D|| 6s \rangle$}
 & \multicolumn{2}{c}{$\langle 6p_{3/2} ||D|| 6s \rangle$}\\
\hline
DHF                     &  1425&4  &  160&94  &  5&2777   &  7&4264\\
\smallskip
SD                      &  2438&0  &  310&71  &  4&4829   &  6&3075\\
\smallskip
$\Delta$(CCSDvT)        &  -136&9  &  -20&92  &  0&0256   &  0&0363 \\
\multicolumn{8}{c}
{Complementary corrections }\\
Line dressing           &   -12&5  &   -2&16  &  0&0094   &  0&0107  \\
Vertex dressing         &     4&3  &    0&29  & -0&0067   & -0&0088  \\
\smallskip
MBPT-IV (core triples,...)
                        &     7&8   &    1&14  &  0&0001   &  0&0001 \\
\smallskip
Breit + QED             &    -6&5\footnotemark[1]
                                   &
\multicolumn{2}{c}{}
                                        &  0&0024 \footnotemark[2]
                                                          &           \\
\smallskip
Extrapol. for $l=\infty$ \footnotemark[3]
                        &     5&0  &    0&37  & -0&004    & -0&006     \\
\smallskip
Final CCSDvT + corrections
                        &  2299&2  &  289&43  &  4&5097   &  6&3398        \\
Experiment              &  2298&2  &  291&89(8) \footnotemark[4]
                                 &  4&5049(17)\footnotemark[5]
                                                          &  6&3404(3)\footnotemark[6] \\
\hline \hline
\end{tabular}
\end{center}
\qquad \qquad \qquad \qquad
\footnotemark[1]{Ref.~\cite{SapChe03}};
\footnotemark[2]{Refs.~\cite{Der00,FlaGin05}};
\footnotemark[3]{Ref.~~\cite{Saf00}};
\footnotemark[4]{Ref.~\cite{RafTan97}};
\footnotemark[5]{Refs.~\cite{RafTan98,BouCruDul06}};
\footnotemark[6]{Ref.~\cite{BouCruDul06}}.
\end{table*}

\begin{acknowledgement}
This work was supported in part by the US National Science
Foundation, by the US NIST precision measurement grant, and by the
Russian Foundation for Basic Research under Grant Nos.\
04-02-16345-a and 05-02-16914-a. This manuscript was
completed with a partial support by NSF through a grant for the
Institute for Theoretical Atomic, Molecular, and Optical Physics at
Harvard University and Smithsonian Astrophysical Observatory.
\end{acknowledgement}

\bibliographystyle{epj}

\begin{thebibliography}{58}

\bibitem{LeeYan56}
T.D. Lee, C.N. Yang, Phys. Rev. \textbf{104}, 254 (1956)

\bibitem{WuAmbHay57}
C.S. Wu, E.~Ambler, R.W. Hayward, D.D. Hoppes, R.P. Hudson, Phys.
Rev.
  \textbf{105}, 1413 (1957)

\bibitem{Zel59}
{\rm Ya}.B. Zel'dovich, Zh. Eksp. Teor. Fiz. \textbf{36}, 964
(1959), [Sov.
  Phys.-JETP {\bf 9} 681, (1959)]

\bibitem{BouBou74}
M.A. Bouchiat, C.~Bouchiat, J. Phys. \textbf{35}, 899 (1974)

\bibitem{BarZol78}
L.M. Barkov, M.S. Zolotorev, Pis'ma Zh. Eksp. Teor. Fiz.
\textbf{27}, 379
  (1978), [JETP \ Lett. {\bf 27}, 357 (1978)]

\bibitem{WooBenCho97}
C.S. Wood, S.C. Bennett, D.~Cho, B.P. Masterson, J.L. Roberts, C.E.
Tanner,
  C.E. Wieman, Science \textbf{275}, 1759 (1997)

\bibitem{BouBou97}
M.A. Bouchiat, C.~Bouchiat, Rep.\, Prog.\, Phys. \textbf{60}, 1351
(1997)

\bibitem{GinFla04}
{\rm J.S.M}.~Ginges, V.V. Flambaum, Phys. Rep. \textbf{397}, 63
(2004)

\bibitem{Khr91}
I.B. Khriplovich, \emph{Parity non-conservation in atomic phenomena}
(Gordon
  and Breach, New York, 1991)

\bibitem{Ram99}
M.J. Ramsey-Musolf, Phys.\ Rev.\ C \textbf{60}, 015501 (1999)

\bibitem{AntArnArr05}
P.L. Anthony, R.G. Arnold, C.~Arroyo, K.~Bega, J.~Biesiada, P.E.
Bosted,
  G.~Bower, J.~Cahoon, R.~Carr, G.D. Cates et~al., Phys. Rev. Lett.
  \textbf{95}, 081601 (2005)

\bibitem{ZelMcFAda02}
G.P. Zeller, K.S. McFarland, T.~Adams, A.~Alton, S.~Avvakumov,
L.~de~Barbaro,
  P.~de~Barbaro, R.H. Bernstein, A.~Bodek, T.~Bolton et~al., Phys. Rev. Lett.
  \textbf{88}, 091802 (2002)

\bibitem{QweakWWW}
Homepage of QWeak collaboration at Jefferson Lab,
www.jlab.org/qweak/

\bibitem{DzuFlaSus89}
V.A. Dzuba, V.V. Flambaum, O.P. Sushkov, Phys.\ Lett.\ A
\textbf{141}, 147
  (1989)

\bibitem{BluJohSap90}
S.A. Blundell, W.R. Johnson, J.~Sapirstein, Phys.\ Rev.\ Lett.
\textbf{65},
  1411 (1990), {P}hys.\ Rev.\ D {\bf 45}, 1602 (1992)

\bibitem{BenWie99}
S.C. Bennett, C.E. Wieman, Phys.\ Rev.\ Lett. \textbf{82}, 2484
(1999)

\bibitem{CasCurDom99}
R.~Casalbuoni, S.~{de Curtis}, D.~Dominici, R.~Gatto, Phys.\ Lett.\
B
  \textbf{460}, 135 (1999)

\bibitem{Ros00}
J.L. Rosner, Phys.\ Rev.\ D \textbf{61}, 016006 (2000)

\bibitem{Ros02}
J.L. Rosner, Phys. Rev. D \textbf{65}, 073026 (2002)

\bibitem{Der00}
A.~Derevianko, Phys.\ Rev.\ Lett. \textbf{85}, 1618 (2000)

\bibitem{DzuHarJoh01}
V.A. Dzuba, C.~Harabati, W.R. Johnson, M.S. Safronova, Phys.\ Rev.\
A
  \textbf{63}, 044103 (2001)

\bibitem{KozPorTup01}
M.G. Kozlov, S.G. Porsev, I.I. Tupitsyn, Phys.\ Rev.\ Lett.
\textbf{86}, 3260
  (2001)

\bibitem{JohBedSof01}
W.R. Johnson, I.~Bednyakov, G.~Soff, Phys.\ Rev.\ Lett. \textbf{87},
233001
  (2001)

\bibitem{KucFla02}
M.~Kuchiev, V.~Flambaum, Phys.\ Rev.\ Lett. \textbf{89}, 283002
(2002)

\bibitem{MilSusTer02}
A.I. Milstein, O.P. Sushkov, I.S. Terekhov, Phys.\ Rev.\ Lett.
\textbf{89},
  283003 (2002)

\bibitem{MilSusTer03a}
A.~Milstein, O.~Sushkov, I.~Terekhov, Phys.\ Rev.\ A \textbf{67},
62103 (2003)

\bibitem{SapPacVei03}
J.~Sapirstein, K.~Pachucki, A.~Veitia, K.T. Cheng, Phys.\ Rev.\ A
\textbf{67},
  052110 (2003)

\bibitem{FlaGin05}
V.V. Flambaum, {\rm J.S.M}.~Ginges, Phys. Rev. A \textbf{72}, 052115
(2005)

\bibitem{ShaPacTup05}
V.M. Shabaev, K.~Pachucki, I.I. Tupitsyn, V.A. Yerokhin, Phys. Rev.
Lett.
  \textbf{94}, 213002 (2005)

\bibitem{Der02}
A.~Derevianko, Phys.\ Rev.\ A \textbf{65}, 012106 (2002)

\bibitem{DzuFlaGin02}
V.A. Dzuba, V.V. Flambaum, {\rm J.S.M}.~Ginges, Phys. Rev. D
\textbf{66},
  076013 (2002)

\bibitem{GueChaJac03}
J.~Guena, D.~Chauvat, P.~Jacquier, E.~Jahier, M.~Lintz,
S.~Sanguinetti,
  A.~Wasan, M.A. Bouchiat, A.V. Papoyan, D.~Sarkisyan, Phys. Rev. Lett.
  \textbf{90}, 143001 (2003)

\bibitem{GueLinBou05}
J.~Guena, M.~Lintz, M.A. Bouchiat, Phys. Rev. A \textbf{71}, 042108
(2005)

\bibitem{For93}
N.~Fortson, Phys. Rev. Lett. \textbf{70}, 2383 (1993)

\bibitem{KoeSchNag03}
T.W. Koerber, M.H. Schacht, W.~Nagourney, E.N. Fortson, J. Phys. B
\textbf{36},
  637 (2003)

\bibitem{VetMeeMaj95}
P.A. Vetter, D.M. Meekhof, P.K. Majumder, S.K. Lamoreaux, E.N.
Fortson, Phys. \
  Rev. \ Lett. \textbf{74}, 2658 (1995)

\bibitem{GomOroSpr06}
E.~Gomez, L.A. Orozco, G.D. Sprouse, Rep. Prog. Phys. \textbf{69},
79 (2006)

\bibitem{AmiDulGut02}
C.~Amiot, O.~Dulieu, R.F. Gutterres, F.~Masnou-Seeuws, Phys. Rev. A
  \textbf{66}, 052506 (2002)

\bibitem{BouCruDul06}
N.~Bouloufa, A.~Crubellier, O.~Dulieu, (private communications)

\bibitem{CoeKum60}
F.~Coester, H.G. K\"{u}mmel, Nucl.\ Phys. \textbf{17}, 477 (1960)

\bibitem{Ciz66}
J.~\v{C}\`{i}\v{z}ek, J. Chem.\ Phys. \textbf{45}, 4256 (1966)

\bibitem{BisKum87}
R.F. Bishop, H.G. K\"{u}mmel, Physics Today \textbf{3}, 52 (1987)

\bibitem{BluJohLiu89}
S.A. Blundell, W.R. Johnson, Z.W. Liu, J.~Sapirstein, Phys.\ Rev.\ A
  \textbf{40}, 2233 (1989)

\bibitem{BluJohSap91}
S.A. Blundell, W.R. Johnson, J.~Sapirstein, Phys.\ Rev.\ A
\textbf{43}, 3407
  (1991)

\bibitem{EliKalIsh94}
E.~Eliav, U.~Kaldor, Y.~Ishikawa, Phys.\ Rev.\ A \textbf{50}, 1121
(1994)

\bibitem{AvgBec98}
E.N. Avgoustoglou, D.R. Beck, Phys.\ Rev.\ A \textbf{57}, 4286
(1998)

\bibitem{SafDerJoh98}
M.S. Safronova, A.~Derevianko, W.R. Johnson, Phys.\ Rev.\ A
\textbf{58}, 1016
  (1998)

\bibitem{SafJohDer99}
M.S. Safronova, W.R. Johnson, A.~Derevianko, Phys.\ Rev.\ A
\textbf{60}, 4476
  (1999)

\bibitem{LinMor86}
I.~Lindgren, J.~Morrison, \emph{Atomic Many--Body Theory}, 2nd~edn.
  (Springer--Verlag, Berlin, 1986)

\bibitem{DerEmm02}
A.~Derevianko, E.D. Emmons, Phys.\ Rev.\ A \textbf{66}, 012503
(2002)

\bibitem{CanDer04}
C.C. Cannon, A.~Derevianko, Phys. Rev. A \textbf{69}, 030502(R)
(2004)

\bibitem{DerPor05}
A.~Derevianko, S.G. Porsev, Phys. Rev. A \textbf{71}, 032509 (2005)

\bibitem{PorDer06Na}
S.G. Porsev, A.~Derevianko, Phys.\ Rev.\ A \textbf{73}, 012501
(2006)

\bibitem{Saf00}
M.S. Safronova, Ph.D. thesis, University of Notre Dame (2000)

\bibitem{Moo58}
C.E. Moore, \emph{Atomic energy levels}, Vol. III (National Bureau
of
  Standards, Washington, D.C., 1958)

\bibitem{SapChe03}
J.~Sapirstein, K.T. Cheng, Phys. Rev. A \textbf{67}, 022512 (2003)

\bibitem{RafTan97}
R.J. Rafac, C.E. Tanner, Phys. Rev. A \textbf{56}, 1027 (1997)

\bibitem{RafTan98}
R.J. Rafac, C.E. Tanner, Phys. Rev. A \textbf{58}, 1087 (1998)

\end{thebibliography}
%

\end{document}